# Atomic-scale observations of electrical and mechanical manipulation of topological polar flux-closure

Xiaomei Li[1,2,7,‡], Congbing Tan[3,4,‡], Peng Gao[2,5,*], Yuanwei Sun[2], Pan Chen[1,7], Mingqiang Li[2], Lei Liao[1,7], Ruixue Zhu[2], Jinbin Wang[4], Yanchong Zhao[1,7], Lifen Wang[1], Zhi Xu[1,7,8], Kaihui Liu[5,6], Xiangli Zhong[4,*], and Xuedong Bai[1,7,8,*]

[1]*Beijing National Laboratory for Condensed Matter Physics and Institute of Physics, Chinese Academy of Sciences, Beijing 100190, China*

[2]*International Center for Quantum Materials, and Electron Microscopy Laboratory, School of Physics, Peking University, Beijing 100871, China*

[3]*Department of Physics and Electronic Science, Hunan University of Science and Technology, Xiangtan 411201, China*

[4]*School of Materials Science and Engineering, Xiangtan University, Hunan Xiangtan 411105, China.*

[5]*Collaborative Innovation Centre of Quantum Matter, Peking University, Beijing 100871, China*

[6]*State Key Laboratory for Artificial Microstructure & Mesoscopic Physics, School of Physics, Peking University, Beijing 100871, China*

[7]*School of Physical Sciences, University of Chinese Academy of Sciences, Beijing 100049, China*

[8]*Songshan Lake Materials Laboratory, Dongguan, Guangdong 523808, China*

[‡]*These authors contributed equally to this work.*

E-mails: *p-gao@pku.edu.cn; xlzhong@xtu.edu.cn; xdbai@iphy.ac.cn*

**Abstract**

The ability to controllably manipulate the complex topological polar configurations, such as polar flux-closure via external stimuli, enables many applications in electromechanical devices and nanoelectronics including high-density information storage. Here, by using the atomically resolved *in situ* scanning transmission electron microscopy, we find that a polar flux-closure structure in PbTiO$_3$/SrTiO$_3$ superlattices films can be reversibly switched to ordinary mono ferroelectric *c* domain or *a* domain under electric field or stress. Specifically, the electric field initially drives the flux-closure move and breaks them to form intermediate *a/c* striped domains, while the mechanical stress firstly starts to squeeze the flux-closures to convert into small vortices at the interface and form a continues dipole wave. After the removal of the external stimuli, the flux-closure structure spontaneously returns. Our atomic study provides valuable insights into understanding the lattice-charge interactions and the competing interactions balance in these complex topological structures. Such reversible switching between the flux-closure and ordinary ferroelectric domains also provides the foundation for applications such as memories and sensors.

**Introduction**

Topological structures in ferromagnetic materials (e.g. skyrmions) have shown great potential applications in electromechanical devices[1], spintronic information storage devices[2,3] and logic devices[4-6] due to their topologically protected states[7]. Given the similarities between ferromagnetism and ferroelectricity, the complex low-dimensional topological polar structures that inherent to ferroelectric materials, such as polar flux-closure[8,9] and vortex[10,11], have also been attracting increasing attentions in recent years[1,8,9,11-15]. Polar flux-closure is a stable topological domain structure formed by the interplay of charge, orbital, and lattice degrees of freedom, with head-to-tail continuous electric dipoles[8,16,17]. The small size of the flux-closure array, which are more preferable for information storage applications due to their smaller size compared to ferromagnetic topological structures, have the potential to stimulate the development of ultrahigh-density (~$12\times10^{12}$ bits per square inch[18]) memory devices.

Most of their applications require the ability to manipulate the polar states of these topological structures through the external stimuli. Considering that the formation of flux-closure is a result of the delicate competition balance between electrostatic and strain boundary conditions, the phase transition from topological configuration to other polar states is expected to occur under external stimuli which can disrupts this balance. In fact, theoretical investigations have predicted that the polar topological structure can be converted to ordinary ferroelectric phases driven by temperature, electric field, or stress [13,18-20]. For instance, the occurrence of a transformation from the vortex to ordinary ferroelectric phase has been predicted in $Pb(Zr_{0.5}Ti_{0.5})O_3$ nanoparticles to accommodate the homogeneous electric field[12]. On the other hand, experimental investigations have successfully observed the transition of domain topology at mesoscale by cooling through the Curie temperature[21,22] or applying an electric field with piezoresponse force microscopy (PFM)[12,23-25]. However, the surface probe-based PFM is limited by spatial resolution thus cannot afford us the ability to obtain full information regarding the nanosized polar topological structures, which have highly structural inhomogeneity and strong charge-lattice coupling, present in the buried thin film. In contrast, the recent advances in atomically resolved *in situ* scanning transmission electron microscopy (STEM) have demonstrated the ability to capture the structural evolutions in individual polar vortex[26,27], thus, motivating the present study that utilizing such a cutting-edge technique to investigate whether or not the topological polar flux-closure can be controllably switched to ordinary ferroelectric domains and further reveal the transition behaviors at the atomic level.

Here we experimentally demonstrate that under the electric field, the topological flux-closure in the $PbTiO_3$ (PTO) /$SrTiO_3$ (STO) superlattice film firstly transforms into an intermediate *a/c* striped domain by gradually moving, breaking the flux-closure structures and eventually form mono *c* domain. Meanwhile, the compressive stress tends to squeeze the flux-closure to convert into small vortices accompanied with formation of dipole wave before finally transforming into mono *a* domain. In both cases, flux-closures returns to its original state when the external stimuli are removed. In other words, flux-closure as a topological structure can be reversibly broken and recovered by either an electric field or a compressive strain. Such an ability to reversibly switch between the flux-closure and ordinary ferroelectric domain structures enables many possible applications for nanoelectronics and electromechanical devices.

**Results**

A dark-field transmission electron microscopy (TEM) image (Fig. 1a) shows the morphology of the (PTO (8 nm) /STO (4.8 nm))$_8$ film grown on the GdScO$_3$ (110) ((001)$_{pc}$, where pc indicates pseudocubic indices) substrate with an SrRuO$_3$ buffered-layer by pulse laser deposition (PLD), where the subscript "8" denotes eight layers of the PTO/STO unit. In this image, the STO layers show a bright and uniform contrast, while the PTO layers adopt a two dimensions (2D) periodic sinusoidal array with dark contrast (*c* domains, with the c axis along the out-of-plane direction), whereas the domains with triangular configurations give rise to another periodic array with bright contrast (*a* domains, with the c axis along the in-plane direction)[8]. This morphology is identical to what has been previously reported for the alternate clockwise and counterclockwise flux-closure quadrants[8,9]. A high-angle annular dark field (HAADF) scanning TEM (STEM) image (Fig. 1b) shows that the contrast in each PTO layer is essentially uniform, indicating that the chemical compositions are homogeneous. Slight variation of the contrast indicates the existence of domain walls. The domain structure can be mapped using geometric phase analysis (GPA)[28,29]. The red periodic wave-like structure, as shown in Fig. 1c, represents the *c* domains, while the remaining blue-colored area denotes the *a* domains in the PTO layers and the STO layer is uniformly dark blue, which is consistent with contrast in the dark field image. Therefore, both dark field and the atomically resolved STEM images can be utilized to monitor the evolution of flux-closure under external stimuli. The former has better temporal resolution while the latter has better spatial resolution.

Selected area electron diffraction (SAED) patterns of the film are shown in Fig. 1d. The diffraction spots were enlarged at the top right, corresponding to the $(100)_{pc}$ spot, and the surrounding satellite diffraction spots originated from the ordered flux-closure and the superlattice periods. The typical area (labeled as the blue frame in Fig. 1b and c) is magnified in Fig. 1e, which shows the atomically resolved HAADF-STEM images overlaid with displacement vectors representing the relative magnitude and direction of the spontaneous polarization[30,31], showing a typical flux-closure pattern. The adjacent *c* domains with opposite polarization directions give rise to the 180° domain wall, at the bottom of which, the small *a* domain is the product of a strong depolarization field. However, the formation of large *a* domain is mainly to balance the large tensile strain[16]. It is worth noting that the larger *a* domains correspond to the triangular regions with bright contrast in the dark field TEM images and the triangular shaped regions in blue in the GPA mappings. The 90° and 180° domain walls are demarcated by yellow and white dashed lines, respectively. Periodic repetition of structure, shown in Fig. 1e, forms the alternate clockwise and counterclockwise flux-closure, which is periodic in both the horizontal and vertical directions. The corresponding lattice *c* mapping (Fig. 1f) obtained by a two-dimensional Gaussian algorithm[31,32] from Fig. 1e gives rise to the sinusoidal feature of the flux-closure, indicating that the flux-closure is closely associated with the c parameter, which reflects the strain distribution and effectively distinguish the domain structure.

To investigate the structural evolution, the cross-section (PTO (8 nm) /STO (4.8 nm))$_8$ film was then subjected to *in situ* electrical and mechanical stimuli via a scanning tungsten probe in a TEM with various image and diffraction modes. Electric fields were applied along out-of-plane direction, providing a driving force for the growth of the *c* domains. A typical chronological HAADF-STEM image series (Fig. 2a) and the corresponding GPA images (Fig. 2b) show that nucleation of phase transition occurs under the tip (the eighth layer); as the contrast in GPA becomes uniform red, which is a feature of the *c* domains at +5 V, and the second layer transforms into tilted stripes, which correspond to *a/c* domains at +7 V. With an increase in the voltage, more layers show the *c* domain characteristics, indicating a transformation from the flux to the *c* domain intermediated with *a/c* domain mixture.

In order to display the evolution process for each flux-closure, the GPA corresponding to the representative areas with different voltages shown in Fig. 2a labeled with the yellow, orange and blue outlines were enlarged, as depicted in Fig. 2c-e. The specific evolution from flux-closure to the *a/c* domains are denoted by the black circle in Fig. 2c, in which the *c* domain with upward polarization (pointing to right, $c^+$) gradually shrinks via motion of 90° and 180° domain walls as well as the of flux-closure core and eventually disappears. Another region in Fig. 2d shows the *a* domain is gradually converted into *c* domain with downward polarization (pointing to left, $c^-$). The inhomogeneous switching is likely related with the inhomogeneous distribution of electric fields due the tip-shape electrodes (see details in Supplementary Fig. 1). Near the tip at the top surface with larger strength of field in Fig. 2e, a relatively large $c^-$ domain is generated. Furthermore, the critical electric field strength to initiate the transition from the flux-closure to the *a/c* domain was estimated to be 40 MV/m. Under negative voltage, the dark field images show the same phenomenon (Supplementary Fig. 2), indicating the flux-closure can be controllably switching to $c^+$ (upward) and $c^-$ (downward) via electric fields.

The transition is reversible. Removal of the electric fields leads to the spontaneous recovery from the *c* domain to the flux-closure state. In Fig. 3a, the atomically resolved HAADF images of the tilted stripes taken at + 9 V show an alternating array of *a/c* domains, which are confirmed by the corresponding lattice c mapping shown in Fig. 3b. As the voltage decreases, the representative, indicated by the black ellipse, display the evolution from the *a/c* domains to the flux-closure; that is, the red colored *c* domain appears in the originally blue-colored *a* domain and gradually becomes larger until it joins with an adjacent *c* domain, resulting in its transformation into the flux-closure structure (Fig. 3b). Note that the new $c^+$ domain nucleates at the corner of *a* domain, $c^-$ domain and interface of PTO/STO (highlighted by the black ellipse in Fig. 3b), and a new flux-closure is formed accompanied with nucleation of $c^+$ domain. Once the $c^+$ domain reaches the other interface (corner), another flux-closure is generated.

Additionally, the flux-closure can also be manipulated by a strain field. Application of a compressive force along the film typically favors in-plane polarization, leading to the growth of the large *a* domain, such process is chronologically shown in Fig. 4a and the corresponding GPA is shown in Fig. 4b. The compressive stress in the first five images was controlled to gradually increase, and gradually decrease in the sixth image. Flux-closure transition driven by mechanical stress begins at the region in contact with the tip with relatively large strength. As the applied mechanical force increases, the transformed region further expands toward the substrate, and more flux-closures are converted into *a* domains. Enlarged magnifications in Fig. 4c,d show that the red stripes that correspond to the *c* domains gradually shrink while the blue regions corresponding to large *a* domain expand. These large *a* domains become larger and larger and connect each to form wave-like shape, so called dipole wave in previous phase field simulations. Meanwhile, the red regions that correspond to vortex cores become discrete and highly located near the interface. In fact, such dipole wave and interfacial vortices have been predicted by phase field simulations[26,33]. To further illustrate the conversion of the flux-closure under the action of compressive stress, the lattice c mapping of the transitional boundary region was depicted in Supplementary Fig. 3, which show a sharp 90° interface between the formed a-domain and pristine flux-closure phase. Again, removal of the external stress leads to the spontaneous recovery from *a* domain to the flux-closure structure. As shown in the sixth image of Fig. 4c, when the stress is reduced, the flux-closure structure is restored to its original position. The process of stress-induced flux-closure transition and recovery can also be confirmed using the continuously recorded SAED patterns and dark field images, as shown in Supplementary Figs. 4 and 5. The stress field for phase transition is estimated to be 15 µN, as shown in Fig. 4e.

Therefore, the flux-closure can be controllably and reversibly manipulated by electric and stress fields, as schematically illustrated in Fig. 5. Under electric field, the unfavored *c* domain starts to shrink via 90° and 180° domain walls motion. Since in this system the 90° domain walls are always rooted at the flux-closure core, during shrinkage the elimination of unfavored *c* domain naturally leads to the elimination of 90° domain walls and destruction of the flux-closure, eventually forming ordinary *a/c* domain stripes. During the recovery from *a/c* strips to flux-closure, the new *c* domain (previously unfavored) nucleates at the interface corner and a new flux-closure is formed simultaneously. The new *c* domain continues to expand to reach the other interface to generate another flux-closure. Notably, the mechanical induced phase transition is distinct. Rather than the generation of *a/c* domain under electrical field, the stress tends to firstly push the flux-closure cores toward to interface and squeeze them to become vortices. Meanwhile the *c* domains are shrinking with polarization rotation ~90° to emerge with the *a* domain to form a dipole wave. When further be stressed, these vortices become smaller and smaller and finally disappear along with the generation of a mono *a* domain. In both cases, the removal of the external fields leads to spontaneous recovery to the original flux-closure structure, demonstrating the reversibility.

**Discussion**

Flux-closure domains, as one of the topological defect structures, can be transitioned under external stimuli. Under electrical field, the switching of unfavored *c* domain is initiated with the motion of domain walls and the flux-closure core rather than the break of the flux-closure core, indicating the flux-closure core indeed is relatively stable. The recovery process that is initiated with the formation of flux-closure and tip-like *c* domain shape with inclined and charged domain walls (in Fig.3b), further confirms the stability of flux-closure. Under mechanical stress, each flux-closure converts to a vortex with the disappear of 180° domain walls and sustains for long time until all the 90° domain walls and *c* domain are eliminated. All these phenomena evidence that the relatively high stability of topological polar structures compared to ordinary ferroelectric domain walls.

At last, we compare the observed switching dynamics in polar flux-closure with previous reported results from polar vortex. The phase-field simulation has demonstrated the ability to switch the polar vortex, i.e., with the application of an electrical field, the vortex cores with opposite curls move closer until reach the same lateral position then produces new *a* domains, and the reversible back-switching take place when removal the applied field[34]. Experimentally, it is reported that the vortices first move toward each other and then move away, finally it becomes a stable *c* domain which doesn't recover spontaneously[26]. Under mechanical stress, the polar vortex can be switched to *a* domain and reversible recovers after removal of stress[27]. In our study, the switching behaviors of flux-closure is even more complicate than that of the vortex, because the transformation of flux-closure involves not only the topological core but also the ordinary ferroelectric domain walls. During switching of a flux-closure under both electric and mechanical stimuli, the topological core indeed seems more stable than the domain walls. Nevertheless, for both polar vortex and flux-closure, the required external stimuli (both electric field and stress field) for phase transition, is comparable with that to switch the ordinary domains[30,35,36], indicating the practicability to fabricate flux-closure-based nanoelectronics and electromechanical devices.

In summary, we controllably switch the topological flux-closure to ordinary ferroelectric domains by using applied electrical and mechanical stress. The atomically resolved *in situ* (S)TEM technique enables us to capture all the details during transition. For electrical loading, the topological structure starts to gradually break the flux-closures and convert into intermediate a *a/c* domains at ~40 MV/m. Eventually a mono *c* domain can be obtained, of which the direction of the polarization direction depends on the direction of the external electric field. For mechanical loading, the compression stress (~15 μN) results in the shrinkage of the *c* domain in the flux-closure accompanied with the conversion to small polar vortices. The large *a* domain expands to form a continues dipole wave with small polar vortices buried at the interface. With further being stressed, a mono *a* domain can be obtained. These switching processes are completely reversible. The ability to reversibly manipulate the topological polar structure and controllably switching them to ordinary ferroelectric domain structures provides a solid foundation for its application in nanoelectronics and electromechanical devices.

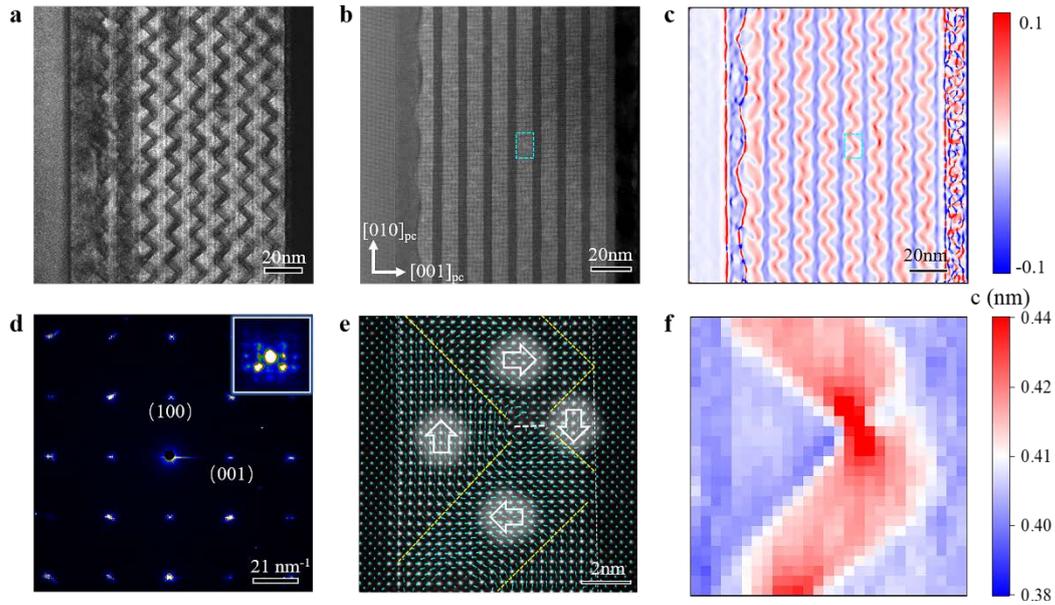

**Fig. 1 | Characterization of the flux-closure in (SrTiO₃)/(PbTiO₃) superlattices. a,** Cross-sectional dark-field TEM image of a (SrTiO₃)/(PbTiO₃) formed by reflection g = 200$_{pc}$, showing the alternative arrangement of SrTiO₃ and PbTiO₃ on GdScO₃ substrates. Flux-closure domains show wave-like shape. **b,** Atomically resolved HAADF-STEM image. **c,** Geometric Phase Analysis (GPA) analysis of the corresponding STEM image shows the distribution of out-of-plane strain $\mathcal{E}_{yy}$. **d,** A selected area electron diffraction (SAED) pattern for the SrTiO₃/PbTiO₃ film. Enlarged (100)$_{pc}$ spots on the top right showing the satellite diffraction spots from the ordered flux-closure and superlattice periods. **e,** The enlarged view of HAADF-STEM image corresponding to the region of blue frame in the b and c overlay of the polar displacement vectors displacement vectors denoted by the blue arrows show the flux-closure polar pattern in the PbTiO₃ layer. The yellow and white dashed lines indicate the 90° and 180° domain walls, respectively. White arrows denote the spontaneous polarization direction of PTO. **f,** Lattice c mapping corresponding to the image e.

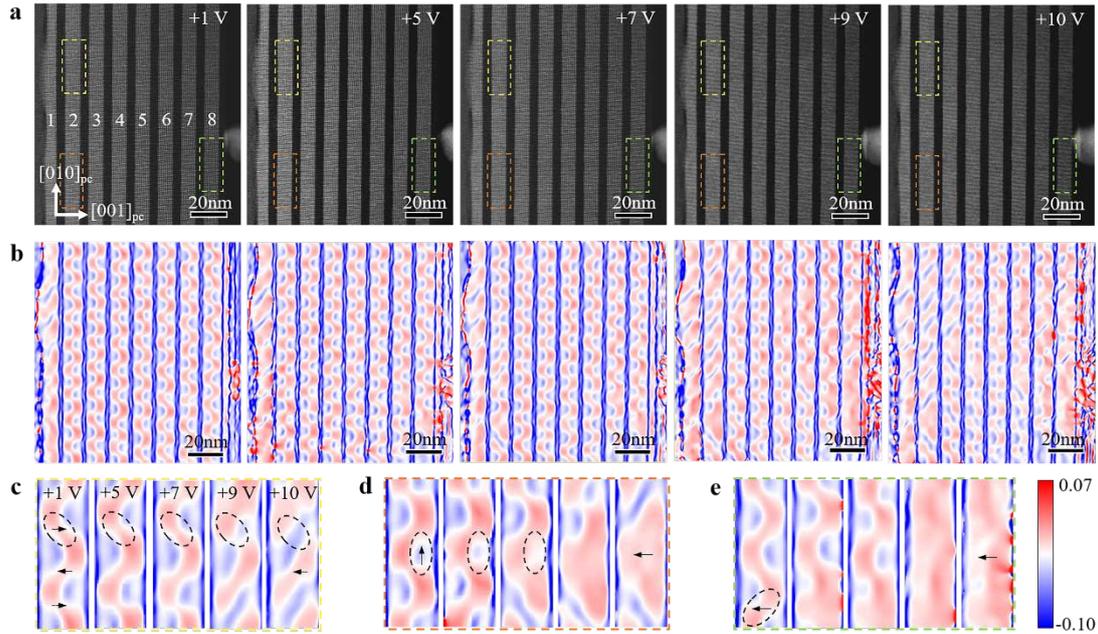

**Fig. 2 | Tracking the transition process of flux-closure under increased external electrical fields at atomic scale. a**, A HAADF-STEM image series acquired during different electrical fields. **b**, The corresponding out-of-plane strain ($\varepsilon_{yy}$) maps extracted from the GPA shows the evolution of domain pattern under increasing electric fields. **c**, **d**, and **e,** Enlarged GPA images of the areas labeled by the yellow, red, and green rectangles in a.

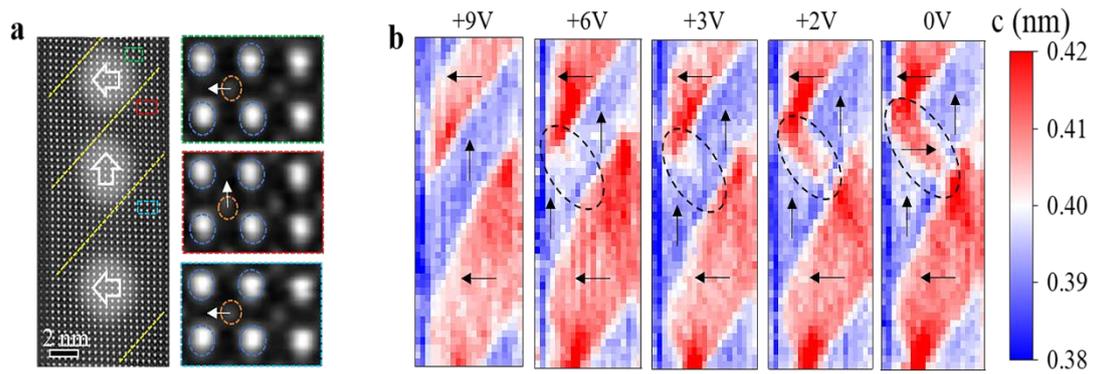

**Fig. 3 | The process of flux-closure recovery. a**, An atomically resolved HAADF-STEM image showing the switched area becoming *a/c* domains. Enlarged HAADF image of the area labeled by the same color shows the displacement of the Ti respective to Pb. The blue and orange circles denote the positions of Pb and Ti columns, respectively. **b**, Lattice c mapping series corresponding to the HAADF-STEM images acquired during gradual removal of the electrical fields, showing the flux-closure recovery process. The dark dashed circle highlights the growth of *c* domain.

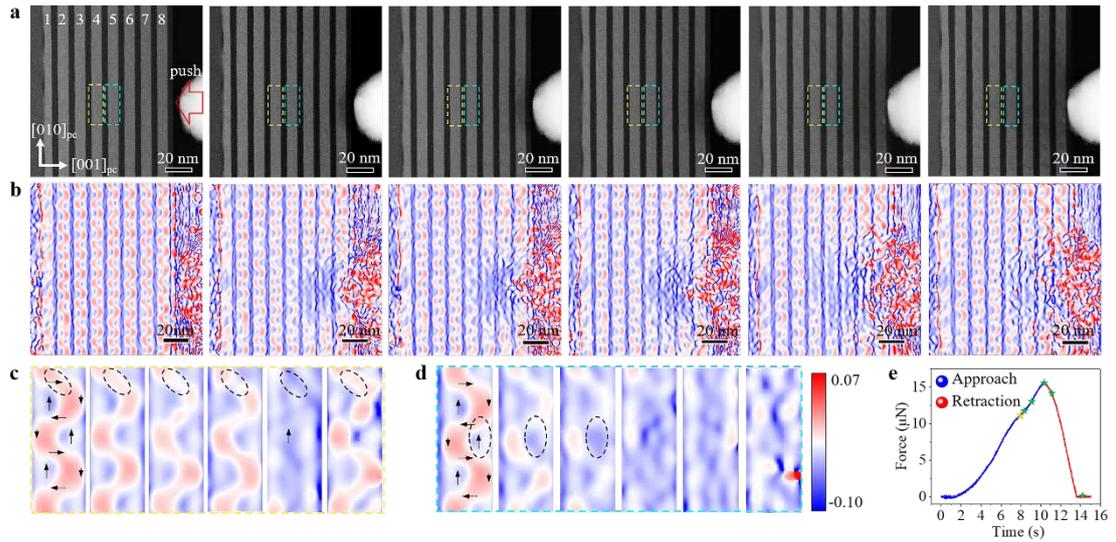

**Fig. 4 | Tracking the process of flux-closure transition under mechanical stress at atomic scale. a**, A HAADF-STEM image series acquired during a mechanical stress load. **b**, The corresponding out-of-plane strain ($\varepsilon_{yy}$) maps extracted via the GPA analysis. **c** and **d,** Enlarged GPA images of the areas labeled by the yellow and blue rectangle in a. **e**, The mechanical load versus time. The blue points representing the approach branch and red points corresponding to the retraction branch.

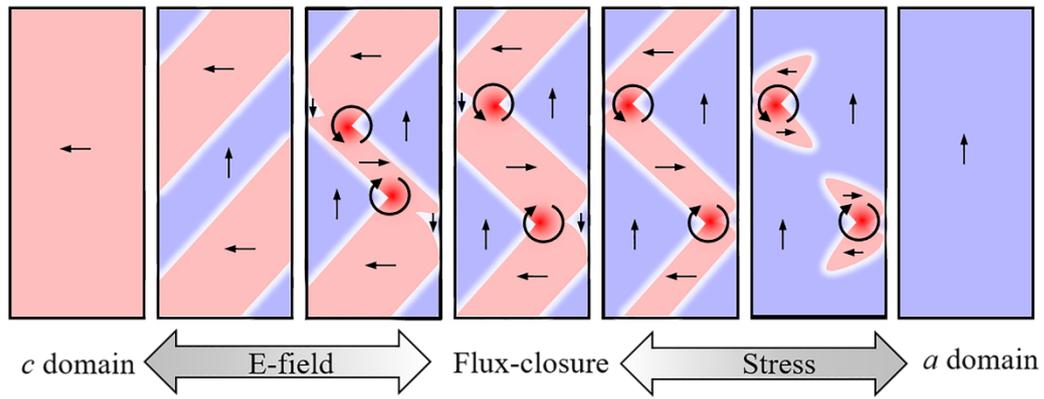

**Fig.5 | Schematic illustration of the flux-closure evolution.** Flux-closure transforms into *a* domain under a compressive strain and *c* domain under an electric field.

**Acknowledgments**

This research was supported by the National Key R&D Program of China (Grants No. 2016YFA0300804); National Equipment Program of China (ZDYZ2015-1); the National Natural Science Foundation of China (Grants No. 11974023, 51672007, 11875229, and 51872251); the Key R&D Program of Guangdong Province (2018B030327001, 2018B010109009); Bureau of Industry and Information Technology of Shenzhen (No. 201901161512); and the "2011 Program" Peking-Tsinghua-IOP Collaborative Innovation Center for Quantum Matter.

**Authors contributions**

P.G., C.B.T., and X.D.B. conceived the idea; C.B.T. grew the samples assisted by X.L.Z. and J.B.W.; X.M.L., Y.W.S., M.Q.L., and P.C. performed the electron microscopy experiments and data analysis under the direction of P.G.; L.L. simulated electric field; Z.X. contributed *in situ* TEM holders; X.M.L and R.X.Z. wrote the paper under the direction of P.G.; X.D.B. supervised the entire project. All authors contributed to the work through fruitful discussion and/or comments to the manuscript.

**Competing interests**

The authors declare that they have no competing interests.

**Data availability**

All data needed to evaluate the conclusions in the paper are present in the paper and/or the Supplementary Materials. Additional data related to this paper may be requested from the corresponding authors.

**Methods**

**Thin Film Growth.** Superlattice film of (PTO (8 nm) /STO (4.8 nm))$_8$ were deposited on the (110)-GdScO$_3$ substrate with an SRO buffered-layer in a pulsed laser deposition (PLD) system (PVD-5000) equipped with a KrF excimer laser ($\lambda$ = 248 nm). Ceramic targets of SrRuO$_3$, Pb$_{1.1}$TiO$_3$ (10 mol% excessive amount of lead to compensate the evaporation loss of Pb) and SrTiO$_3$ were used for the SRO layer and PTO/STO superlattices deposition. The SRO-buffered layer was first grown at 690 °C and 80-mTorr oxygen pressure, and then the substrate was cooled to 600 °C for the deposition of the PTO/STO superlattice at a 200-mTorr oxygen pressure. Selecting the appropriate laser energy was crucial for ensuring the layer-by-layer growth of the SRO, PTO, and STO sublayers, and here, the SRO and PTO/STO superlattices were prepared under laser energies of 390 and 350 mJ/pulse, respectively. By controlling the growth time, thicknesses of the PTO and STO layers were held at 8 nm and 4.8 nm. Right after growth, the superlattice films were cooled down to room temperature at 50 °C min$^{-1}$ cooling rates.

**TEM Sample Preparation.** Cross-sectional TEM samples were prepared by a conventional method that includes mechanical polishing and then ion beam milling. The ion-beam milling was carried out using argon ion milling (Gatan 695) with the acceleration voltage of 2.6 kV and reduced to 0.1 kV to reduce the irradiation damaged layers.

**STEM Characterization.** High-angle annular dark field (HAADF) images for character the sample were carried out in an aberration-corrected JEOL Grand ARM 300 CFEG and FEI Titan Cube Themis G2. The collection semi-angles snap for the HAADF imaging ranges from 54 to 220 mrad (JEOL) and 48 to 200 mrad (Titan).

***In situ* (S)TEM.** *In situ* (S)TEM experiment was carried out using an aberration-corrected JEOL ARM 300F at 300kV in STEM mode and TEM mode. The real-time diffraction pattern, dark field TEM images, and the atomically HAADF-STEM images were recorded in JEOL ARM 300F with a double-tilt holder provided by ZEPTools Technology Company. The quantitative measurement of the mechanical loads was performed in FEI F20 microscope operated at 200 kV in TEM mode together with Hysitron system (PI 95).

**The Electric Field Calculations.**

We calculated the distribution of electrical fields in STO/PTO superlattice using Ansoft Maxwell software based on finite element analysis and Maxwell's equations[36-38]. For simplicity, all calculations assume SRO and tungsten as perfect conductors, and STO/PTO superlattice as perfect insulator. We estimated the tip contact width to be ~20 nm. The relative dielectric constant of STO and PTO is assumed to be 300 and 200. The model for calculation consists of 8 superlattice periods, in which the thickness of PTO and STO is estimated to be ~8 nm and ~4.8 nm, respectively. The in-plane and out-of-plane component of electrical field distribution are determined by calculating electric potential distribution and its partial derivatives.